\documentstyle[preprint,natbib]{aastex}
\input epsf

\begin{document}

\title{Tidal Interactions and Disruptions of Giant Planets on Highly
Eccentric Orbits}
\author{Joshua A.\ Faber, Frederic A.\ Rasio, and Bart Willems}
\affil{Department of Physics and Astronomy, Northwestern University}

\begin{abstract}

We calculate the evolution of planets undergoing a strong tidal
encounter using smoothed particle hydrodynamics (SPH), for a range of
periastron separations.  We find that outside the Roche limit, the
evolution of the planet is well-described by the standard model of
linear, non-radial, adiabatic oscillations.  If the planet passes
within the Roche limit at periastron, however, mass can be stripped
from it, but in no case do we find enough energy transferred to the
planet to lead to complete disruption.      
In light of the three new extrasolar planets discovered 
with periods shorter than two days, we argue that the
shortest-period cases observed in the 
period-mass relation may be explained by a
model whereby planets undergo strong tidal encounters with stars,
after either being scattered by dynamical interactions into highly
eccentric orbits, or tidally captured from nearly parabolic orbits.
Although this scenario does provide a natural explanation for the
edge found for planets at twice the Roche limit, it does not explain
how such planets will survive the inevitable expansion that results
from energy injection during tidal circularization.
\end{abstract}

\section{Introduction}

Approximately 17\% of extrasolar planets discovered to date
can be classified as ``hot Jupiters,'' gas giant
planets in very tight orbits (with orbital periods of $< 10$ days)
around solar-like stars \citep{Sas,May,Bou,Kon}.  Of these 21 planets,
17 have periods of less than five days.   
In this context, strong tidal interactions between a
giant planet and its central star have become an important problem. Many
studies
have focused on understanding the mechanisms and consequences of
tidal {\em dissipation\/} in these systems \citep{RTLL,LTL,FRS,GLB,
Sas,IP2,OL,IP1}. Here instead we examine {\em
dynamical\/}
interactions, in which a giant planet on a nearly-parabolic orbit passes
very close to
the central star. There are at least two scenarios where such interactions
would occur.
The first is the scattering scenario for explaining the high eccentricities
of extrasolar planets \citep{RF,WM96,LI97,FHR,MW02,AL03}. 
In this scenario, protoplanetary systems containing several
giant planets of comparable masses become
dynamically unstable, leading to strong scattering between planets. Planets
scattered inward may sometimes undergo strong tidal interactions with the
central star, perhaps even leading to capture onto a much shorter-period
circular orbit. In the second scenario,
discussed recently by Gaudi \citeyearpar{Gaudi}, 
``free-floating'' planets in the dense
cluster environments where most stars are formed would be tidally captured
by protostars, in a
manner reminiscent of the old tidal-capture scenario for forming compact
binaries in
globular clusters \citep{FPR,PT,LO}.
As far as the close encounter is concerned, the only difference between
these two scenarios
is whether the nearly-parabolic orbit of the incoming planet is in fact
slightly hyperbolic
or elliptic. 

\begin{figure}
\centering \leavevmode \epsfxsize=6in \epsfbox{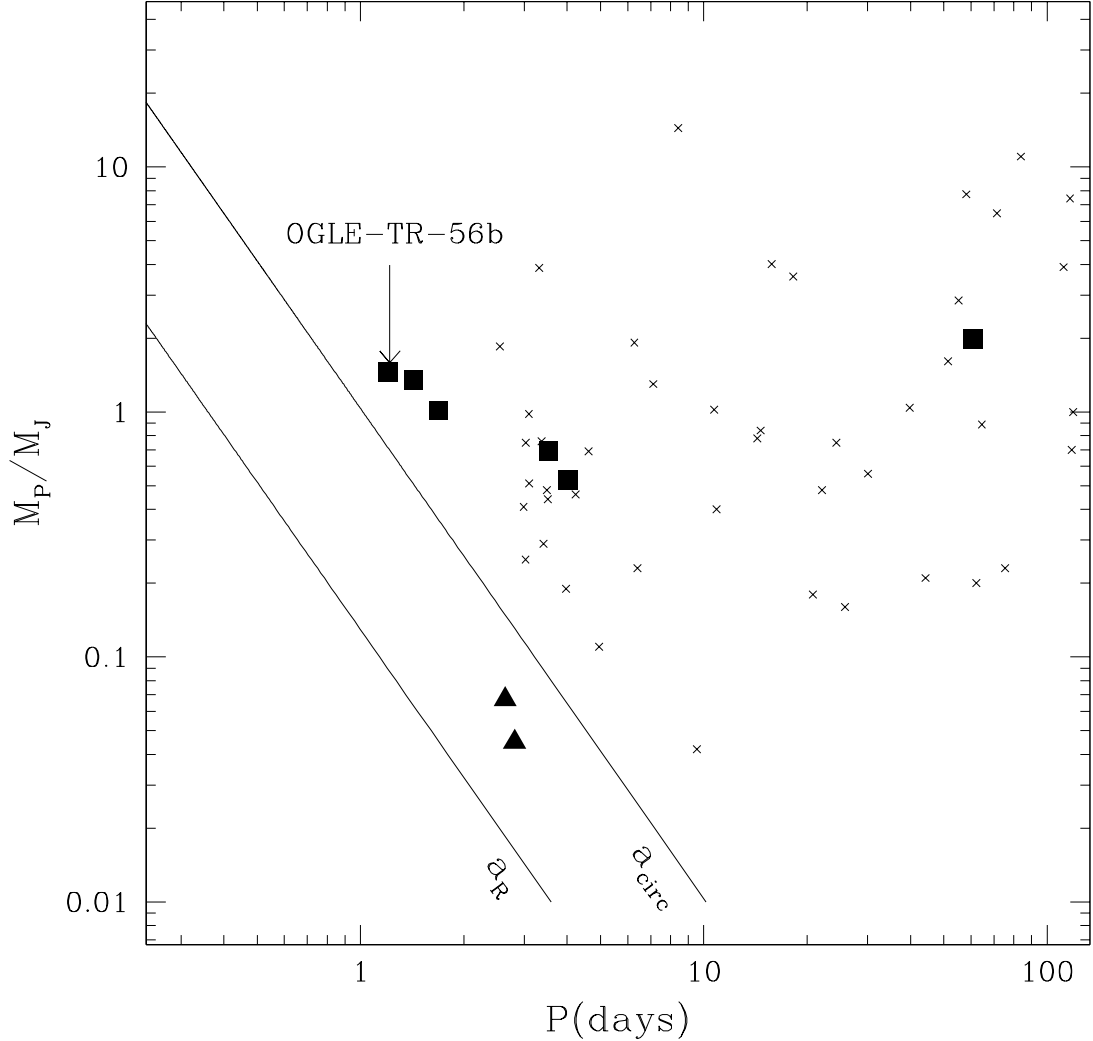}
\caption{Minimum mass $M_P \sin i$ versus orbital period for the
  current observed sample of planetary companions.  OGLE-TR-56b is one of
only six planets whose orbital inclination is known, all of which
  are marked by squares (for these, we show the actual mass).
In all six cases, since the inclination is
  determined from eclipses, $i\ge 80^\circ$.
Triangles represent the possibly lower-mass ``hot Neptunes'', GJ436 b and 55 Cnc e, which
  may have a qualitatively different structure than the more massive
  planets on the figure.  The Roche
  limit, $a_R$, defined via Eq.~\protect\ref{eq:roche},
  is shown for a planet with a radius equal to that of Jupiter,
as is the ideal circularization radius, $a_{circ}$, defined as an orbit with a
  separation twice as large as the Roche limit. Data are taken from the
  extra-solar planets catalog at {\tt http://www.obspm.fr/encycl/cat1.html}.} 
\label{fig:census}
\end{figure}

In this paper we study the strong tidal interaction between a giant planet
and a 
solar-like star using 3-D numerical hydrodynamic calculations. We focus on
encounters with periastron
separations of a few solar radii, which can lead to significant dissipation
of orbital energy and mass loss. Since the details of the interaction
are insensitive to the sign of the total energy, our results can be easily
applied to the case of highly-eccentric elliptic orbits or low-energy
hyperbolic orbits. The main questions we address concern the final fate of
the planet
following an interaction, and the possibility of an observational signature
of strong tidal interactions. Surprisingly,
we find that complete disruption of a Jupiter-like planet {\em outside\/} a
solar-like star
is never possible. If the interaction is close enough for the planet to be
stripped of
a significant fraction of its mass, its orbit always {\em gains\/} enough
energy to become unbound (even if the initial orbit was bound). Even closer
interactions would lead
instead to a physical collision with the star. Observationally, hot Jupiters
are confined
to a region of parameter space that appears to follow closely a simple
definition of
the ``tidal limit'', shown in Fig.~\ref{fig:census}. Data points
indicated by squares represent planets whose masses are known, since
the mass function $M_p \sin i$ is constrained by the observation of
transits.  In all other cases, the mass shown represents a lower
limit.  The only planets that may fall within this limit, shown as
triangles, have a much smaller mass, close to that of Neptune, and may
well be structurally different from typical gas giants, as we discuss below.

It is not immediately clear that any common definitions of the tidal limit should be relevant here. Indeed, each is made under
assumptions that are violated for highly eccentric planetary orbits,
e.g., circular orbits and synchronized spins.  We note one crucial fact
about all the tidal limits discussed below, however.  They all have
the same physical scalings, with different coefficients, since the
underlying dimensional analysis is the same for each.

The Hill radius, $r_H$, is defined in the context of the restricted
three-body problem, and is commonly used in describing
the orbits of a planet's satellites.  Based on simple point-mass
mechanics, it is found that a satellite can orbit stably around a
planet of mass $M_P$ in a circular orbit of radius 
$a$ around a star of mass $M_*$ 
so long as its own orbit has a semimajor axis less than
\begin{equation}
r_H=a\left(\frac{M_P}{3(M_P+M_*)}\right)^{1/3}\approx 0.69\, q^{1/3}a,
\end{equation}
where the latter relation holds for small mass ratios, i.e.,
$q=M_P/M_* \ll 1$.
We note that what Gaudi \citeyearpar{Gaudi} refers to as the ``Roche
limit'' is found by a brief calculation to be the Hill radius instead.

The Roche lobe radius, $r_R$, is defined in terms of the classical
stellar two-body problem.  For two point masses in a circular 
orbit, there exists a
critical equipotential surface in the rotating frame around each body, within
which all corotating fluid is bound to it.  This volume can be used
to define a characteristic volume-averaged 
radius; for mass ratios much different than
unity the Roche lobe around the less massive body
is roughly spherical, with a cusp at the inner
Lagrange point.  In the limit of extremely small mass ratios ($q\ll
1$), the Roche lobe radius depends weakly on the compressibility of
the less massive object (the primary is always assumed to be a point
mass; see \citealt{LRS} for an extended discussion).
In most pioneering works (see, e.g., \citealt{Jeans}), it was assumed that the
secondary was completely incompressible (corresponding to
$n=0$, or equivalently, $\Gamma\rightarrow\infty$).
For this case, it was found that
\begin{equation}\label{eq:roche0}
r_R=0.407\, q^{1/3} a.
\end{equation}
Later, \citet{Pac} considered the opposite limit, treating the
secondary as a point-mass (corresponding to the infinitely {\it
  compressible} case).  Based on tabulated results, he found that the 
Roche lobe radius is given by
\begin{equation}\label{eq:roche}
r_R=0.462\, q^{1/3} a,
\end{equation}
and is almost exactly two-thirds the extent of the Hill radius.
This is the definition of the Roche lobe radius that appears in
\citet{Sas}, among many other sources, and will be the
one used throughout this paper, for reasons we will explain in
detail in Sec.~\ref{sec:calc}.
A related quantity we refer to regularly is the Roche limit $a_R$, defined
as the critical separation where the planet fills its Roche lobe; it
can be defined implicitly through
the relation $r_R(a_R)\equiv R_P$.

Remarkably, we find that the present location of the tidal ``edge'' observed
in Fig.~\ref{fig:census} 
would be naturally explained if all planets, with the possible
> exception of the ``hot Neptunes'' shown as triangles,  had been
initially on highly eccentric orbits
and later circularized {\em without significant mass or orbital 
angular momentum loss} at a distance
approximately twice that of the Roche limit, which we will refer to as
the ``ideal circularization radius''.  
Indeed, any initial orbit with extremely high eccentricity
has a (specific) total angular momentum satisfying
\begin{equation}
j^2\simeq 2GM r_p
\end{equation}
where $M$ is the total mass of the system and $r_p$ the periastron
separation of the initial orbit.  The final circular orbit, on the
other hand, satisfies the condition
\begin{equation}
j^2=GM a.
\end{equation}
Assuming that tidal circularization occurs through dissipation of
orbital energy but with no net loss of mass or angular momentum, 
and neglecting spin angular momentum, we conclude that
the orbit will circularize at a separation $a\simeq 2r_p$.  We show
below why we believe the condition $r_p>r_R$ 
determines whether the planet remains in a bound orbit
after the passage.

Our limits are
placed under the assumption that these ``hot Jupiters'' have radii
not very different from that of Jupiter, $R_J=7.14\times10^9~{\rm cm}$. This is
consistent with current measurements for the three innermost planets found
with OGLE, ($R=1.23\pm0.16 R_J$ for OGLE-TR-56, $R=1.08\pm 0.07 R_J$
for OGLE-TR-113, and $R=1.15^{+0.80}_{-0.13} R_J$ for OGLE-TR-132;
\citealt{Sas,Tor,May,Bou,Kon}), as well as recent theoretical
calculations of the structure of giant planets with extremely
short-period orbits \citep{Burr,Chab}.  If indeed the radii of some of these
planets are slightly larger than that of Jupiter, as appears to be the
case for HD 209458 with $R=1.43\pm 0.04 R_J$, our conclusions
remain unchanged, since the location of the ``circularization
separation'' will move slightly to the right on our plot but maintain the same
functional form.  
 
The ``hot Neptunes'', GJ 436 b \citep{butler} and 55 Cnc e
\citep{mcart} are significantly less massive than the other planets
with periods $P<3$~days, assuming they do not have improbably small
inclination angles.  The radii of these planets, however, are likely
to be smaller than that of Jupiter, since an extended envelope would
be blown away by radiation from the parent M dwarf; it has been
suggested that these planets may in fact be composed completely or in
part of rock and ice \citep{butler} and may be the rocky remnant cores of
gas giants which have lost significant amounts of mass to tidal heating
or some other process \citep{mcart}.

The scenario we investigate here differs in its predictions from those
involving the slow inspiral of giant planets all the way to the tidal limit
(as in many popular ''migration scenarios;'' see, e.g.,
\citealt{Trill,GLB}).
This long-term inspiral is expected to produce an ``edge'' 
{\it at} the Roche limit, not at a separation twice as large.
In addition, it is unclear what mechanism would halt the orbital inspiral
before the onset of Roche lobe overflow and mass loss from planets
experiencing radial expansion from tidal heating.  Various
possibilities have been proposed which rely on the evacuation of the
inner protoplanetary disk \citep{KuLe} and tidal interactions involving
the host star's own rotation \citep{FRS}, but it is unclear how any
orbital model involving tidal decay would produce the currently 
observed ``edge'' further out.

One major caveat with highly eccentric orbits concerns the survival of
the planet during orbital circularization.  Previous calculations
have indicated that as energy is injected into the planet during
circularization, its radius should expand significantly,
eventually leading to Roche lobe overflow \citep{BLM,BLL,GLB,GBL}.  Of course,
given that these same results were used to argue for a lack of planets
with periods $P \lesssim 3$ days, it is fair to say that uncertainties
still remain as to the evolution of planetary radii and separations
through tidal dissipation and circularization.  Of particular
importance is determining the rate at which a planet can dissipate
tidal energy, and the effect of the energy dissipation rate on the
planet's radius.  This process is complicated, and may depend
sensitively on the rotation rate of the planet relative to the angular
velocity during periastron passage \citep{IP1}.  In any case, while we
consider an examination of this matter to be an important step in
understanding planetary orbital evolution, it is beyond the scope of
this paper.

Our paper is organized as follows.  In Sec.~\ref{sec:code}, we
describe our Lagrangian SPH code, as well as the parameters used in
our calculations.  In Sec.~\ref{sec:calc}, we detail the results of our
calculations, looking in turn at the case where the planet passes
outside of the Roche limit, and then the case where it passes within,
since this is found to be crucial in determining the future evolution
of the planet.  In Sec.~\ref{sec:concl}, we discuss how these results
affect the current picture of the evolution of giant planets, and
discuss further scenarios to which these results may be applied.
 
\section{Method and Approximations}\label{sec:code}

All of our
calculations were done with a modified version of the
{\tt StarCrash} smoothed particle hydrodynamics (SPH) code, available at
{\tt http://www.astro.northwestern.edu/StarCrash/}.
Several previous versions of this code have
been used to study a wide range of hydrodynamic interactions between stars
(see, e.g.,
\citealt{RS1,FR1,LomF}). SPH is a Lagrangian method which treats the
dynamical evolution of a set of finite-sized fluid particles. The density
of the fluid is computed at the position of each particle using an
interpolation
kernel with compact support, extending over a
characteristic ``smoothing length.''  In our
implementation, the smoothing length around each particle varies
in time so as to provide overlap with a nearly constant number of
neighboring
particles.  Hydrodynamic forces are computed using SPH summation
techniques (for a detailed derivation, see \citealt{RS1}), whereas
gravitational forces are calculated using a grid-based FFT convolution
method. Shock heating is treated by evolving the energy equation with
an artificial viscosity prescription from Balsara \citeyearpar{balAV}.

Computing the full hydrodynamic evolution of both the planet and the
star would be very challenging but is in fact unnecessary.
Indeed, a simple order-of-magnitude estimate shows that the vast
majority of the tidal energy extracted from the orbit will be deposited in
the planet
during the close interaction. Following
\citet{FPR}, we expect that for the passage of a planet
of mass $M_P$ and radius $R_P$
by a star with mass $M_*$ and radius $R_*$ with periastron separation
$r_p$, the tidal energy deposited in the
planet and 
in the star are given respectively by
\begin{eqnarray}
\Delta E_p \simeq f_p^2 \frac{G M_*^2 R_P^5}{r_p^6},\\
\Delta E_* \simeq f_*^2 \frac{G M_P^2 R_*^5}{r_p^6}.
\end{eqnarray}
The dimensionless factors $f_p$ and $f_*$ depend
primarily on the ratio of the dynamical (crossing) time to the internal
dynamical
time of each object. Therefore they are mainly a function of the mean
density of each object, and should have comparable values for the planet
and star.  
As a result, the ratio of energy dissipated in the planet to that dissipated
in
the star is 
\begin{equation}
\frac{\Delta E_p}{\Delta E_*}\simeq \left(\frac{M_*}{M_P}\right)^2
\left(\frac{R_P}{R_*}\right)^5\sim 10,
\end{equation}  
for a Jupiter-like giant planet and a solar-type star.
A more detailed analysis based on an expansion over adiabatic non-radial
oscillation modes (see, e.g., \citealt{LO}) yields exactly 
the same energy ratio
dependence on mass for two equal-density objects to lowest order, 
albeit with a more
complicated functional dependence on the periastron separation $r_p$.
Noting these results, we follow the full 3-D hydrodynamic and
thermodynamic evolution of the planet, but treat the star as a simple
point mass, which interacts with the fluid through gravitational
forces only.  Although our code can handle fluid particles on grazing
trajectories, by treating the stellar surface as an absorbing
boundary that captures all SPH particles that pass within $R_*=10
R_p$, such techniques play no role in the calculations described
here.  In no run did any SPH particle fall within that boundary.
Our implementation of stellar point-masses
is similar in many ways to previous Newtonian SPH
treatments of
tidal interactions between stars and black holes (e.g., \citealp{KL}), 
but here we can
ignore many of the details regarding the effective boundary around the
point-mass. Indeed, the matter lost by the planet
and accreted onto the star during extremely close passages has a minimal
effect
on the stellar mass and completely negligible feedback on the evolution
of the orbit.

We define our units in
terms of the parameters of the planet, setting $G=M_P=R_P=1$.
In terms of the mass and radius of Jupiter ($M_J=1.9\times
10^{30}~{\rm g}$ and $R_J=7.15\times10^9~{\rm cm}$), this yields
characteristic time,
velocity, energy, and angular momentum scales of
\begin{eqnarray}
t&=&1698\left(\frac{M_P}{M_J}\right)^{-0.5}\left(\frac{R_P}{R_J}\right)^{1.5
}~{\rm sec}=1,\\
v&=&4.21\times 
10^6\left(\frac{M_P}{M_J}\right)^{0.5}\left(\frac{R_P}{R_J}\right)^{-0.5}~
{\rm cm/s}=1,\\
E&=&3.37\times 10^{43}\left(\frac{M_P}{M_J}\right)^2
\left(\frac{R_P}{R_J}\right)^{-1}~{\rm erg}=1,\\
J&=&5.72\times 
10^{46}\left(\frac{M_P}{M_J}\right)^{1.5}\left(\frac{R_P}{R_J}\right)^{0.5}~
\rm{erg}\cdot{\rm s}=1.
\end{eqnarray}
In all our calculations we fix the mass of the star to be
$M_*=1000~M_P=0.95 (M_P/M_J) M_{\odot}$.  The planet's equation of
state (EOS) is approximated by a $\Gamma=2$ (or equivalently, $n=1$)
polytrope, i.e., the
pressure $P=k\rho^2$, where the entropy constant $k$
is initially set to a fixed value throughout.  This EOS
has been found to approximate very well the bulk properties of
Jupiter-like planets, even though it differs from the ideal gas form
at low densities \citep{Jup1,Jup2}. In particular, it provides a mass-radius
relation
such that radius is independent of mass, agreeing well with detailed
models (see \citealt{Burr2,Burr}, and references therein).  We note
that should the radius of an extrasolar planet be larger, as is
predicted for giant planets during early stages of their evolution
(see, e.g., \citealt{Saumon}), our results would have to be scaled
accordingly.

To model the planet in our calculations,
we place SPH particles of varying mass in an
equally-spaced hexagonal close-packed lattice, with particle masses
set proportional to the polytropic model's density at the appropriate
radius.
In all calculations shown here, $N=48846$ particles are used to
describe the planet.  This number which yields a characteristic
smoothing length $h/R_P=0.05$, has been found in previous calculations
to yield results which typically conserve overall energy terms to
within $\sim 1\%$ \citep{FR3}.
The planetary mass distribution is relaxed for 30 dynamical times to achieve a
stable configuration, and placed into a very high-eccentricity 
elliptical orbit around the
``point-mass'' star at an initial  separation $a_0=200$.  This is
sufficiently distant that the initial tidal perturbation of the 
planet is negligible.  All initial orbits have
apastron separations of
$r_a=10^4$, equivalent to $4.78\,{\rm AU}$ for the parameters
of Jupiter. 
The periastron separation is varied to span a
range of values $15 \le r_p \le 50$.  
For all runs, the planet is initially
non-spinning in the inertial frame, with the velocities of all particles
set equal to that of the planet's center of mass.

\section{Calculations and Results}\label{sec:calc}

The evolution of planets on nearly parabolic orbits is found to be
critically dependent on whether or not the orbital separation passes within
the Roche limit.  We note that although some of the assumptions used
to define the Roche limit do not apply to the situation we consider
here, the use of the term is still appropriate.
Specifically, the Roche limit problem assumes that both bodies are
point-masses, and that they both corotate at an orbital velocity
corresponding to a circular orbit.  Here, the planet is an extended
object, it is irrotational in the inertial frame, and the angular
velocity during the encounter corresponds to a highly elliptical
orbit.  Still, the crucial functional dependence on the physical
parameters of the system remain exactly the same. We will show
with our calculations that the critical separation for the planet to
undergo mass loss corresponds extremely closely with the classically
defined Roche limit, and thus refer to what is formally the dynamical tidal
mass-shedding limit as the Roche limit.

As we demonstrate in Appendix~\ref{sec:roche}, the classical Roche
limit formulae found in \citet{Pac} and \citet{Egg}, 
which treat both components in
the system as point masses, underestimate the tidal limit
separation (the point at which Roche lobe overflow begins)
for our extended, corotating 
polytropic planetary models placed on circular orbits, by no more than $2\%$.
Indeed, where \citet{Pac} finds the critical separation for Roche
lobe overflow to begin at a separation $a_R=21.64$, we find that it
occurs at a separation that falls somewhere in the range $a_R=21.8-22.0$.

Our dynamical calculations indicate that for highly elliptical 
planetary orbits with periastron separations within this limit, 
mass will be stripped from the planet, where we note that here the planet is
assumed to be irrotational in the inertial frame.  
Funneling out through both the
inner and outer Lagrange points in two streams, this mass loss
has a significant impact on the future evolution of the planet as well as
the orbit, since it can dramatically affect the overall energy and
angular momentum budget of the system.  Orbits with
periastron separations outside the Roche limit still induce energy and
angular momentum transfer, but no mass is exchanged between the planet
and the star, as we describe in detail below.

\subsection{Outside the Roche limit}

\begin{figure}
\centering \leavevmode \epsfxsize=6in \epsfbox{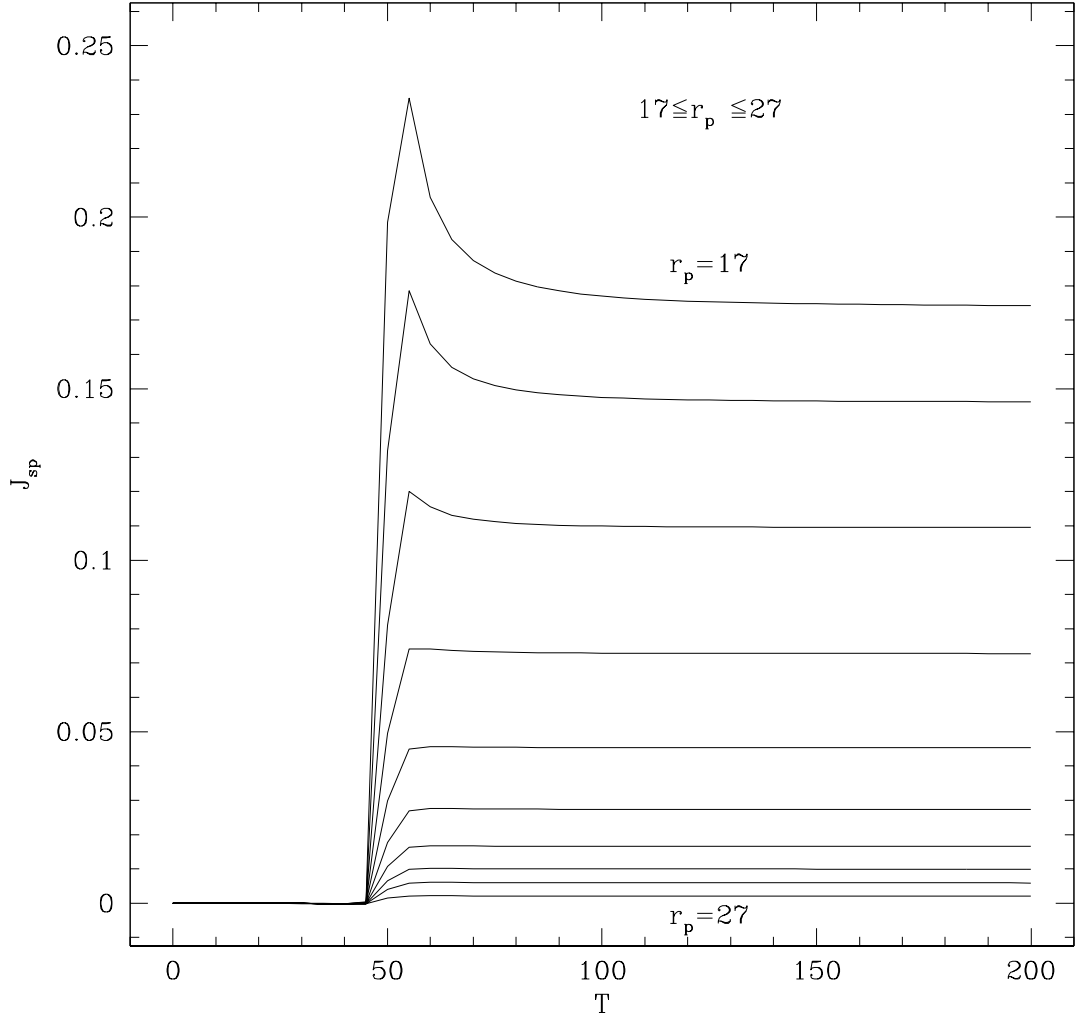}
\caption{The evolution of the spin angular momentum over time for all runs
shown in Table~\protect\ref{table:1} with periastron separations
$17\le r_p \le 27$.  In all cases, we see a spike during the periastron
passage, which lasts from $T=45-50$,
before a slight decrease toward the final relaxed value.  For the
periastron separations shown here, there is a monotonic increase in
the final value of $J_{sp}$ with decreasing $r_p.$}
\label{fig:jspintime}
\end{figure}

\begin{figure}
\centering \leavevmode \epsfxsize=6in \epsfbox{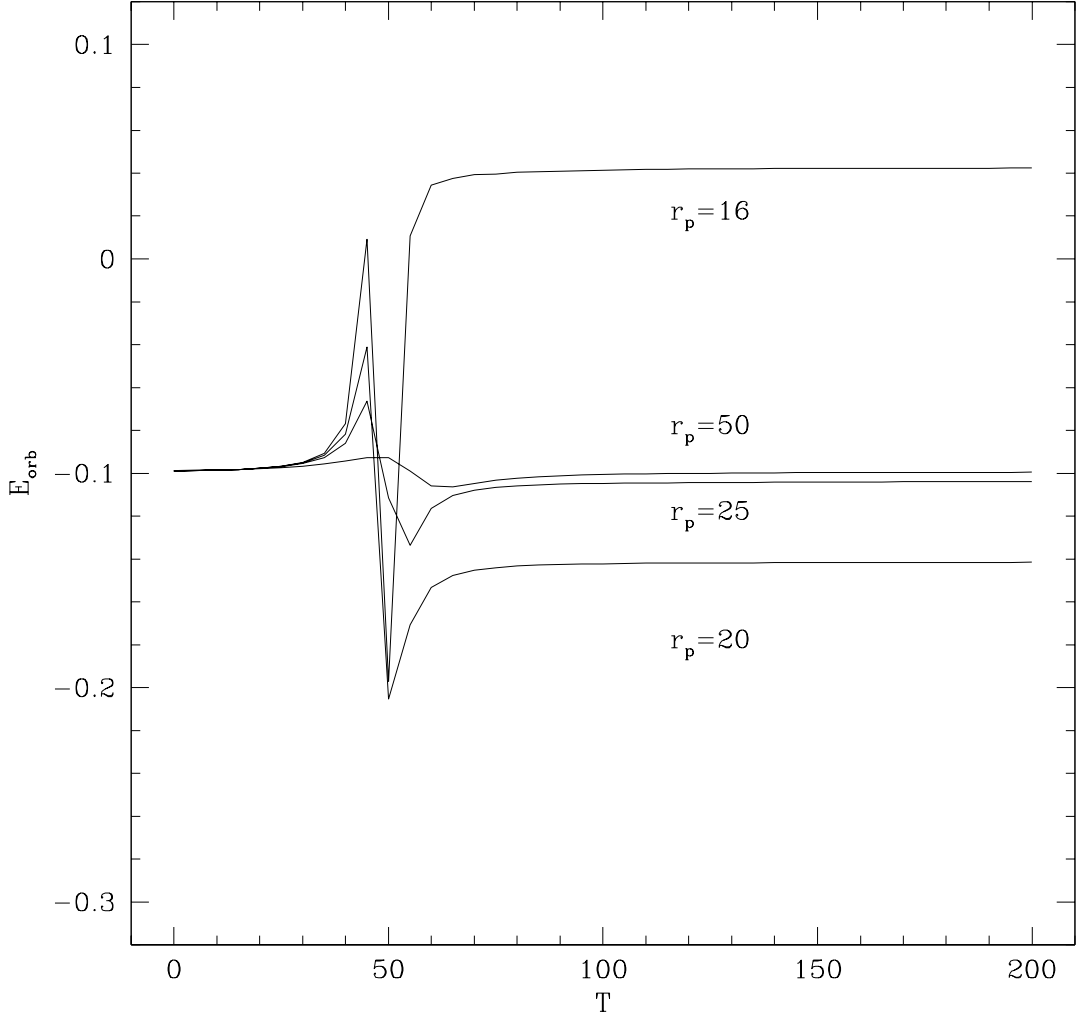}
\caption{The evolution of the orbital energy over time for runs with
$r_p=16$, $20$, $25$, and $50$.  In all cases, we see energy is
injected into the orbit during the first half of the encounter,
followed by a rapid decrease of greater magnitude during the second
half of the encounter, and finally a long period where the total
energy levels off.  For the runs with $r_p\ge 17.5$, the net energy
change is negative, and the final orbit is more bound than the initial
one.  For $r_p=16$, we see that the energy required to strip mass from
the planet leads to the orbit gaining enough energy to unbind completely.}
\label{fig:eorbtime}
\end{figure}

For all systems with initial periastron separations
$r_p \ge 22$, the passage of the planet by the star resulted in a
qualitatively similar pattern of tidal energy and angular momentum transfer.
The temporal behavior of angular momentum and energy transfer is 
demonstrated in
Figs.~\ref{fig:jspintime} and \ref{fig:eorbtime}, for a number of
passages with varying values of $r_p$, and conforms well with the
commonly accepted picture of the tidal interaction process.
In all cases, we see that energy
is transferred into the orbit prior to the periastron passage (which
occurs for all models at $T\simeq 45$), without a noticeable change in
the angular momentum of the planet.  Immediately after periastron,
energy is rapidly transferred via tidal effects into the planet,
with the tidal torque causing a sharp drop in the total
orbital energy.  After the passage, the planet
gradually relaxes toward a new equilibrium spinning configuration.

It should come as no surprise that for the passages outside the Roche limit
in Fig.~\ref{fig:eorbtime}, decreasing the periastron separation leads
to both a decrease in the final orbital energy and an increase in the
spin angular momentum, since the tidal interaction is much
stronger at closer range.  These results cannot be generalized to the
case of orbits passing within the Roche limit,
however, since the final angular momentum and energy distributions are
extremely sensitive to the mass loss that occurs in those cases.

\begin{figure}
\centering \leavevmode \epsfxsize=6in \epsfbox{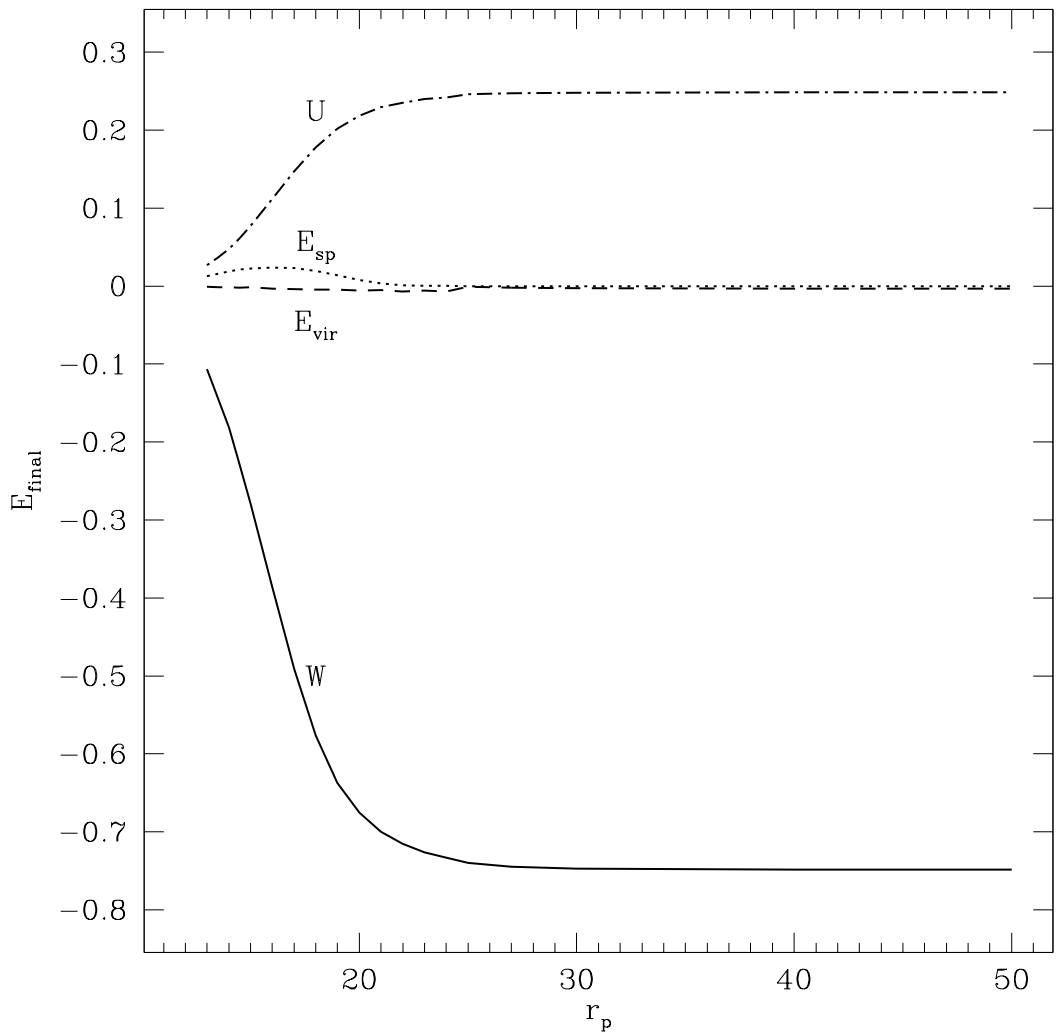}
\caption{The final gravitational binding energy $W$ (solid line), internal
thermodynamic
  energy $U$ (dash-dotted line), and spin kinetic energy $E_{sp}$ (dotted
line)
of the planet, as a function of periastron separation.  Also shown is
  the virial energy $E_{vir}=W+3U+2E_{sp}$, as a dashed line.  We see
  that in all cases the final planetary configuration is nearly
  virialized.  As the total energy for all configurations shown here
  is negative, we conclude that no Jupiter-like planet can be fully
  disrupted by a non-grazing passage past a sunlike star.}
\label{fig:energy}
\end{figure}

As we expect from
simple virial arguments, the tidal heating results in some
degree of expansion, and a less bound structure for the planet, as
shown in  Fig.~\ref{fig:energy}.  This relationship holds for the entire
sequence of orbits we calculated, including those which passed within
the Roche limit, as we discuss in more detail below.
For passages outside the Roche limit, the amount of energy injected into the
planet was not at a sufficient level to unbind any mass from it, down
to our resolution limit (defined by the least
massive SPH particles used near the surface of the planet, 
$m_{min}=1.5\times10^{-7}$).
As the periastron separation decreases toward the Roche
limit, the energy gained by the planet
does increase, causing the planet's radius to increase in corresponding
fashion.  
The various energies and angular momenta describing the final
planetary configurations for all the cases we investigated 
are listed in Table \ref{table:1}.  
For orbits with $r_p \ge 15$ our runs are terminated at $T=200$,
since in all of those calculations the planet had reached a relaxed,
virialized configuration by that point.
For models with smaller periastron separations, we double the
duration of the calculations, letting them run until $T=400$.

\begin{figure}
\centering \leavevmode \epsfxsize=6in \epsfbox{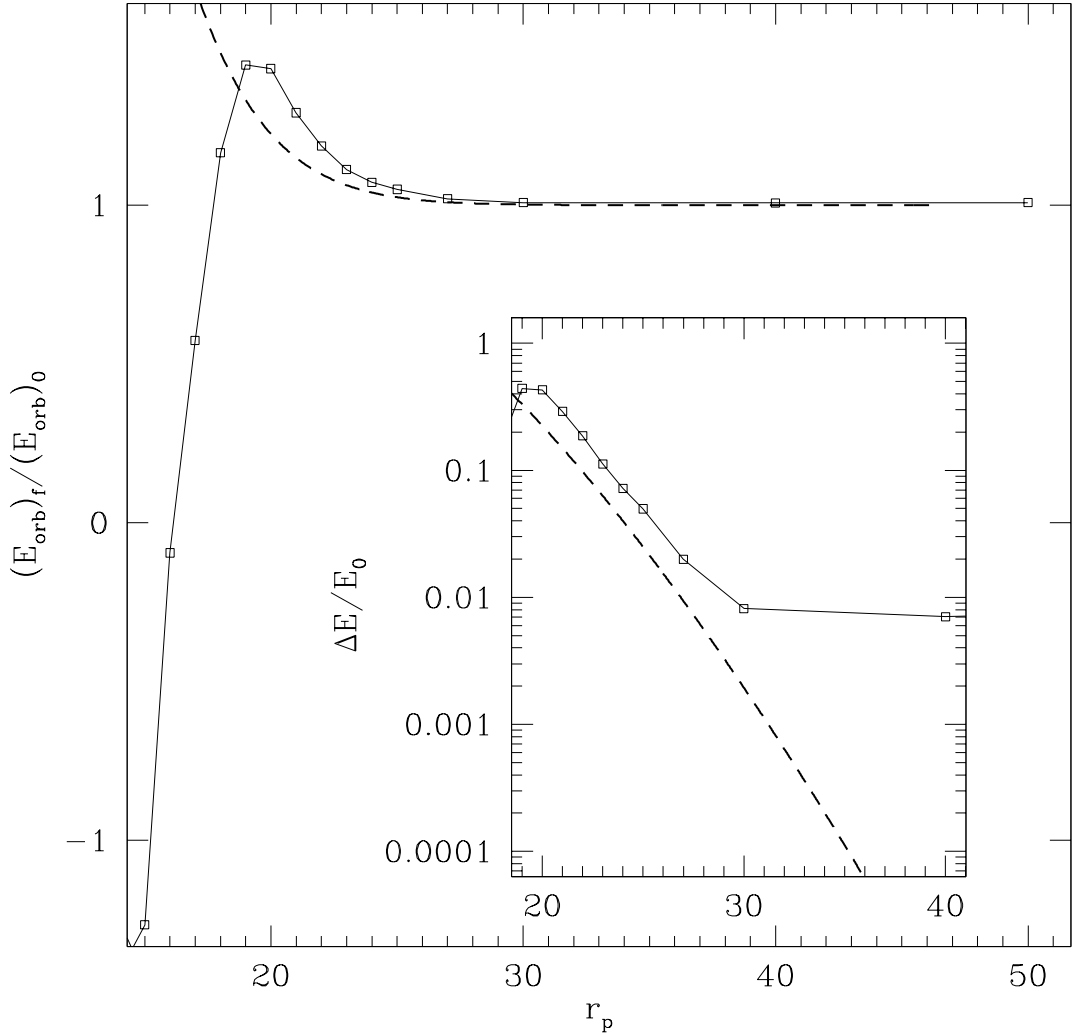}
\caption{The relative change in the orbital energy from the beginning
of our calculations to the end, as a function of the periastron
separation (solid curve).  We
see that this quantity increases as we sweep inward
(becoming more negative), indicating more tightly bound
orbits, until reaching a maximum at $r_p\approx 19.5$.  Within this
periastron separation, systems become successively less bound (the
orbital energy less negative), until at
$r_p\approx 16.2$ we find that the planet becomes unbound from the
star and leaves on a hyperbolic trajectory.  The dashed curve shows
the predicted behavior from a Press-Teukolsky type 
analysis of non-radial adiabatic
oscillations as described in Appendix \protect\ref{sec:leeost}.
This approximation 
scales well at large
separations, up to $r_p\sim 30$ where the systematic errors in the total
energy become larger than the net change, and breaks down for $r_p < 20$
when the linear regime itself is no longer applicable.}
\label{fig:eorb}
\end{figure}

Much of the previous work on tidal capture has used a linear
perturbation formalism, developed by (\citealt{PT}; see also
\citealt{LO}), which treats the fluid response to the 
tidal interaction as a superposition 
of non-radial adiabatic oscillations.  In
Appendix~\ref{sec:leeost}, we summarize the equations describing the
energy loss to tidal perturbations, and give the coefficients
in the expansion for configurations with EOS appropriate for both a gas giant
($\Gamma=2$) and, for completeness, a terrestrial planet ($\Gamma=3$)
\citep{Boss}. 
Outside the Roche limit, we find that the Press-Teukolsky formalism
gives the proper scalings for the tidal interaction process, but
slightly underestimates the overall magnitude of the effect.
In Fig.~\ref{fig:eorb}, we show the final orbital energy as a function
of periastron separation for all the models we computed, as well as
the estimate obtained from the linear
perturbation analysis.
We find that while the power-law scalings are very similar, the
perturbation analysis typically yields a change in the orbital energy
approximately half the magnitude of what we find from our SPH
calculations.  The relationship breaks down completely for
$r_p<20$, when the deformation of the planet undergoing mass loss
clearly becomes nonlinear.
We note that the nearly constant change in orbital energy we find from
our calculations at periastron separations $r_p \ge 30$ are a numerical
artifact, and represent the smallest change in energy we can
accurately measure over the full timescale of one of our evolution
calculations. 

\subsection{Within the Roche Limit}

\begin{figure}
\centering \leavevmode \epsfxsize=6in \epsfbox{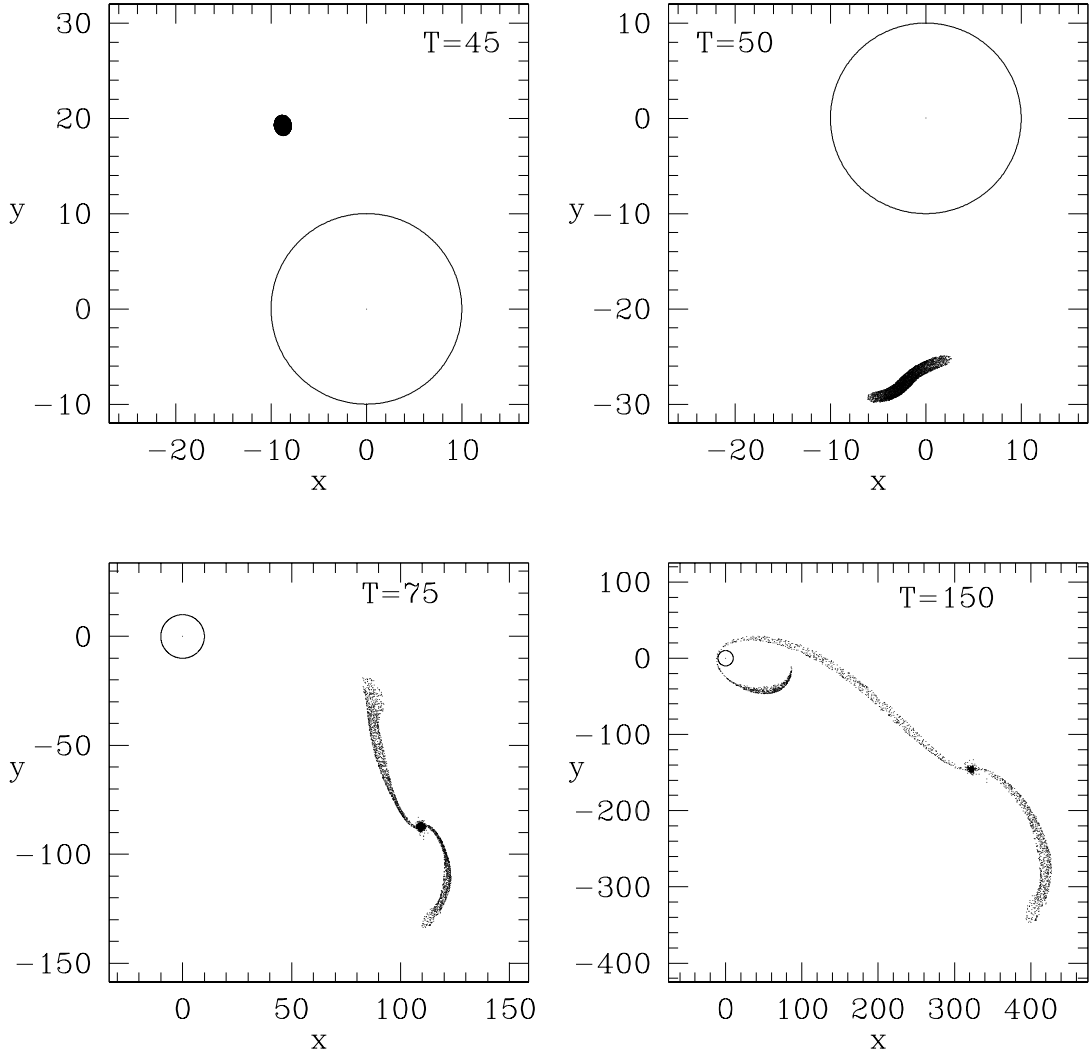}
\caption{Evolution of the planet along an orbit with periastron
separation $r_p=18$.  In the upper left panel, we see the
planet, orbiting counter-clockwise, nearing periastron at $T=45$ past the
star,
whose physical radius is indicated by the circle.  In the
upper-right, at $T=50$, we see the planet immediately after
periastron, with strong tidal effects obvious.  In the lower left, we
see at $T=75$ that a pair of mass-shedding streams have formed, both
toward the star and away.  Eventually, by $T=150$, we see the inner
stream has stretched all the way around the star, as the particles
follow essentially free-fall trajectories in the star's gravitational
well.  Note the different size scales for each plot.  For clarity,
only particles near the orbital plane are shown.}
\label{fig:partplot}
\end{figure}

For initial orbits with $r_p \le 21$, 
the planet passes within the classical Roche limit
(for a mass ratio $q=0.001$, the critical separation
for Roche lobe overflow is $a_R=21.64$, according to Eq.~\ref{eq:roche}).
In all the cases we looked at in this regime, fluid was
stripped from the planet, escaping in extremely narrow streams
through both the inner (L1) and
outer (L2) Lagrange points.
We show the evolution of one such system, with $r_p=18$, in
Fig.~\ref{fig:partplot}.  The axes are defined
such that the planet orbits counter-clockwise along an orbit whose
unperturbed pericenter would fall on the negative x-axis, and the
timescale initialized to the initial configuration at
separation $a_0=200$.   In the first panel, we show the configuration
of the planet at $T=45$, shortly before periastron, as well as the
star, whose physical size is indicated by the large circle. The planet
is beginning to show signs of tidal deformation (looking roughly
ellipsoidal).  Note again
that the stellar size is merely illustrative, and plays no role in the
calculations, since no
matter from the planet crosses within the stellar radius during 
our calculation.  In the
second panel, at $T=50$, immediately after periastron passage, we see
the planet starting to distend further, as matter crosses through the Lagrange
points while tidal energy is transferred into the planet.  At
$T=75$, two mass-shedding streams are clearly evident.  We find that all
particles in the inner stream, representing a total mass $\Delta
m_{in}=0.021$
 are no longer bound to the planet, but remain
gravitationally bound to the star.  By contrast, particles in the
outer stream, representing a total mass $\Delta m_{out}=5.9\times
10^{-3}$,  are bound to neither the star nor the planet, and will be
ejected from the system.  In the final frame, we find at $T=150$ that
fluid in  both streams has assumed an essentially ballistic
trajectory, freely falling in the star's potential well.  Particles from the
inner stream trace out nearly elliptical orbits, retaining enough
angular momentum to pass outside the star's surface, while those in
the outer stream head away from the system on hyperbolic orbits,
leading the path of the planet in an almost cometary fashion.

\begin{figure}
\centering \leavevmode \epsfxsize=6in \epsfbox{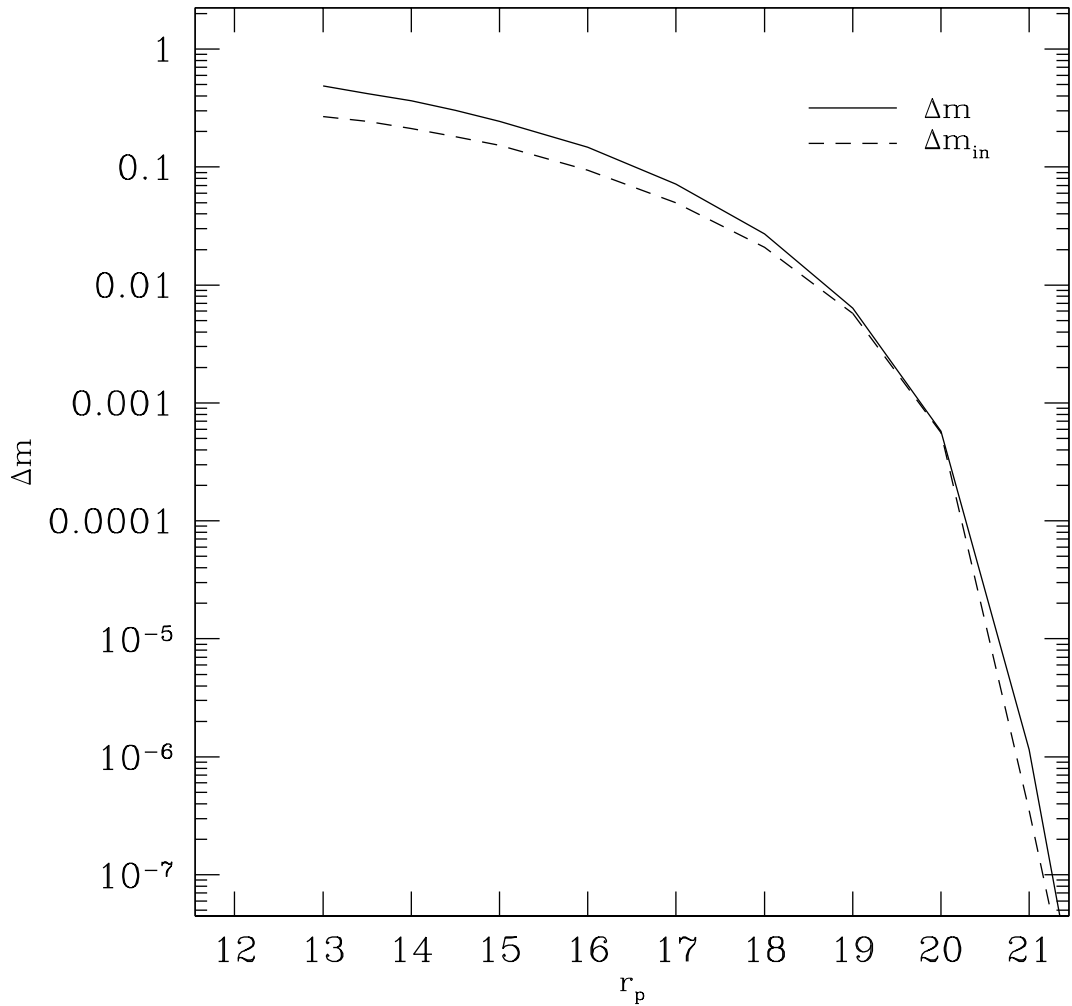}
\caption{The total mass unbound from the planet during the interaction
(solid line) and the mass which ends up bound to the star (dashed line), as a
function of periastron separation.  As a rule, matter in the inner
mass shedding stream ends up bound to the star, whereas matter in the
less massive outer
stream is completely unbound.}
\label{fig:massloss}
\end{figure}

We can make a few general statements with regard to orbits within the
mass-shedding regime.  First, in all cases we investigated, the amount
of mass stripped from the planet
increased with decreasing periastron separation, as shown in
Fig.~\ref{fig:massloss}.
The rise seems to be almost
exponential near the Roche limit, but
flattens out at smaller values, such that even for orbits on which the
planet will
graze the edge of the star during the passage, it will {\it not} be completely
unbound by the interaction.  This is in agreement with previous
results for tidal disruptions of stars around massive black holes
\citep{CL, EK}, that indicate full disruption only occurs for
stars on orbits with periastron separations meeting the criterion 
$\eta \lesssim 1.0$, where the
interaction strength $\eta$ is defined by the relation
\begin{equation}\label{eq:eta}
\eta\equiv\left(\frac{M_2}{M_1+M_2}\right)^{0.5}
\left(\frac{r_p}{R_2}\right)^{1.5},
\end{equation}
where $M_1$ and $M_2$ are the masses of the more massive object and
the body being disrupted, respectively (here, $M_1\equiv M_*$ and
$M_2\equiv M_P$). 
For systems with
$q \ll 1$ and equal-density components, this condition essentially
yields $r_p<R_*$. In other words, planets would to have to pass
within the star in order to be fully disrupted.

In all cases we studied (except the orbit with $r_p=21$, in which
fewer than 10 SPH particles became unbound), the amount of matter
stripped from the planet along the inner stream, which remains bound to the
star, exceeds the amount of matter unbound from the system
through the outer stream.  This asymmetry in the mass of the two
tidal streams is greatest near the Roche limit.  Indeed, for orbits with $18\le
r_p \le 20$, we find that over $75\%$ of the mass stripped from
the planet can be found in the inner stream.
For orbits with smaller periastron separations, especially those
nearing the limit of a grazing collision, the mass ratio in the
two streams nears unity.
While at first glance these results may appear to differ slightly from the picture developed for the
disruption of a star by a massive black hole in \citet{LTH},
\citet{Rees}, and \citet{EK}, we note that 
their calculations were performed for orbits with $\eta\sim 1.0$, 
which represents a periastron separation here of
$r_p=10.0$. Summarized, when the tidal energy $\Delta E$ injected into the
smaller body is greater in magnitude than the binding energy $E_B$ of the
object, we expect it to be disrupted.  The velocities of fluid
elements in the smaller object take on a range of values, depending on
the depth of the passage through the potential well, with a roughly
flat distribution of specific energies centered near zero, since
$\Delta E \gg E_B$ (see Fig.~3 in \citealt{EK}).  The fluid with
negative specific energy becomes bound to the larger body in the
system, and that with positive specific energy unbound, representing
nearly equal amounts.
For the planet-star interactions we investigate here, 
we expect to find equal masses deposited into
the inner and outer streams only in the limit of grazing collisions,
for which $\Delta E\sim E_B$.  Such a conclusion cannot be
generalized to passages with larger values of $r_p$ since the
magnitude of the tidal bulge does not approach the same scale as the
planet's radius, or in other terms, the tidal energy remains smaller
than the overall self-binding energy of the planet.
Furthermore, in terms of the interaction strength $\eta$, we expect
that disruption of a Jupiter-like planet by a solar-type star should
require a tighter passage than for a solar-type star being disrupted
by a massive black hole of $M_{BH}\approx 10^6 M_{\odot}$.  Indeed,
the characteristic expansion velocity $v_{exp}$ of the secondary as it
is being disrupted
scales like $v_{exp}=(M_1/M_2)^{1/6}v_{esc}$, where $v_{esc}$ is the
escape velocity from its surface.  
Thus, the characteristic expansion velocity within a
star being disrupted by a massive black hole ($M_1/M_2=10^6$) is more
than double that for a planet being disrupted by a star
($M_1/M_2=10^3$), relative to the respective escape velocities.  We
conclude that the core of the planet should remain bound for passages
with lower values of $\eta$ than for the stellar-massive black hole case.

For close passages, the planet experiences
a radial expansion, due to strong tidal heating throughout.
As the periastron separation decreases, 
$R_{95}$, defined as the radius enclosing $95\%$ of the planet's
bound mass at the end of the calculation, 
increases to a value a few times larger than that of the
planet prior to the encounter, especially in cases with $r_p \le 18$.
For all mass-shedding systems, the furthest gravitationally bound
particles, at a distance $R_{100}$ from the planet's center of mass, 
were found to be located at $R_{100}>30$.  
This state would almost certainly not
be permanent, and merely reflects the extremely long dynamical relaxation
timescale found in the low-density outer regions of the planet.
Since the dynamical timescale has a power-law dependence proportional
to $\rho^{-0.5}$, the matter furthest from the planet (and thus with
the lowest density) requires considerably more time than we can
feasibly calculate to reach equilibrium.  However, since this
represents an extremely small fraction of the total mass, we don't
expect the long-term relaxation of the planet's outer regions to
affect our results about the dependence of various energy quantities
on the periastron separation.

Several quantities we tabulate do not show monotonic dependence on the
periastron separation of the planet's orbit.  We see in
Fig.~\ref{fig:eorb} that it is only for the range 
$r_p\ge 19.5$ that the final orbital energy becomes more strongly
negative as the periastron separation decreases. For these cases, the
apastron separation of the post-encounter orbit decreases with decreasing  
$r_p$ (the periastron separation, as one should expect, remains
essentially fixed after the encounter, since the specific orbital
angular momentum changes very little during the encounter). 
Indeed, the orbital energy
reaches an extremum and begins to become less negative (the orbit less
tightly bound) as $r_p$
decreases below the critical value $r_p=19.5$.
For orbits with $r_p<17.5$, there is a net {\it gain} in the
orbital energy of the planet.
For periastron separations 
$r_p\lesssim 16.2$, the orbital energy becomes positive, and the planet
leaves the system on an unbound hyperbolic trajectory.  This can also be seen
in Fig.~\ref{fig:eorbtime}, as the run with $r_p=16$ demonstrates a
characteristically different pattern than those at larger
separations.  The tidal interaction leads to a sharp
increase in the orbital energy followed immediately by larger
decrease.  This is followed by
significant mass loss from the planet, leaving only $\sim 40\%$
of the original mass of the planet gravitationally bound.  
A great deal of material which once formed the planet becomes
bound to the star instead, causing a strong {\it increase} in the orbital
energy of the surviving planet.  A similar pattern was
seen for all passages with $r_p \le 16$.

\begin{figure}
\centering \leavevmode \epsfxsize=6in \epsfbox{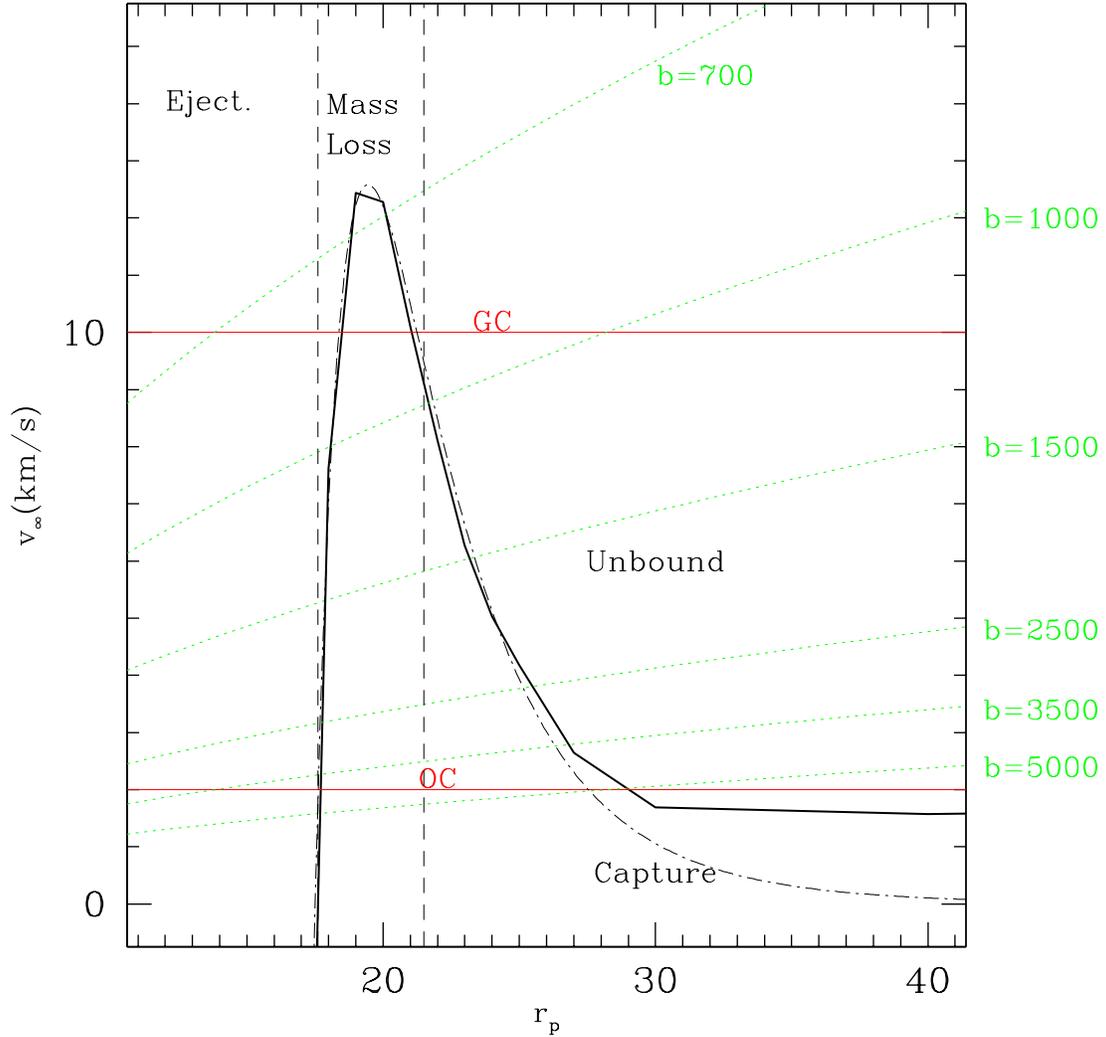}
\caption{The critical value $v_{capt}$ of the relative velocity at large
distances, $v_{\infty}$, for which a planet can be captured,
as a function of the periastron separation (solid line).  Velocities below
the
curve lead to capture, those above the curve to unbound systems.  For
$r_p\lesssim 17.5$, there is a net gain in orbital energy, and no
bound system can be formed. Also shown are curves of constant initial
impact parameter $b$, as well as the typical relative
velocities within a globular cluster ($v_\infty=10~{\rm km/s}$) and an
open cluster ($v_\infty=2~{\rm km/s}$).  The dot-dashed line shows the
approximate fit, Eq.~\protect\ref{eq:vfit}, which we use to estimate
the capture timescale.}
\label{fig:vcapt}
\end{figure}

We can use these results to classify the fate of planets
passing by stars on either bound or unbound orbits.  In Fig.~\ref{fig:vcapt},
we show
the critical relative velocity at large separations leading to
capture, $v_{capt}\equiv \sqrt{2(\Delta
E_{orb})/\mu}$, as a function of $r_p$.  
Here $\mu\approx M_P$ is the reduced mass of the
system.  If the relative velocity of the star and planet at
large separations falls below the critical value, the planet
can be tidally captured during the interaction.
Note that these results may underestimate the true capture
velocity by $\sim 5\%$, since we have ignored tidal dissipation in the
star.  Of course, since the energy loss scales 
$\propto r_p^{-6}$, this represents less than a $2\%$ correction to the
maximum capture radius for a given relative velocity.
We see that in a globular cluster, where the typical relative
velocity is $\sim 10$ km/s, there is a very low probability of forming
bound systems through tidal capture.  For an open cluster, however,
with a typical velocity dispersion of $2$ km/s, 
close passages satisfying $18\lesssim r_p
\lesssim 30$ are likely to meet this criterion. At the high end, our
main source of uncertainty is a
systematic overestimate of $v_{capt}$, since
deviations from equilibrium in our initial
conditions act as a small spurious energy source for the
orbit. 
To confirm our estimate of the maximum capture velocity at large
separations, we computed additional runs with periastron separations of
$r_p=40$ and $r_p=50$, for orbits with zero total energy (parabolic),
and small positive energies (hyperbolic with $v_\infty=0.05=2~{\rm km/s}$
and
$0.1=4~{\rm km/s}$).  As expected, we find that the total change in the orbital
energy is essentially unchanged, since the interaction timescale is
set by the orbital velocity at periastron, which depends very strongly
on the periastron separation but extremely weakly on the total orbital
energy.

To calculate cross sections for these collisions, the
important quantity is the impact parameter of the hyperbolic orbit,
\begin{equation}
b=r_p\sqrt{1+\frac{2GM_*}{r_p v_\infty^2}},
\end{equation}
rather than the periastron distance.  
To give some sense of the size scales involved,
for a globular cluster with
$v_{\infty}=10~{\rm km/s}$, the narrow typical
range of capturable periastron separations, $18.5<r_p<21$,
yields a correspondingly narrow range of impact parameters, 
$770<b<820$. For an open cluster with $v_\infty=2.0~{\rm
km/s}$, the range of periastron separations $18<r_p<30$ corresponds to
a wider range of impact parameters for capture,
$4000<b<5000$, or roughly $2.0~{\rm AU}<b<2.4~{\rm AU}$.

It is relatively straightforward to expand this analysis toward a more
physically realistic picture of tidal capture.  Following
\citet{Gaudi}, we will assume that planets and stars move within some
form of cluster (either globular or open) with the same characteristic
velocity dispersion $\sigma$.  Such a condition would theoretically result
after planets are liberated from their original parent stars by a
series of weak encounters, but before thermal relaxation causes them
to gain sufficient velocity to escape the cluster.  Furthermore, we will
assume that the number density of stars $\nu$ is uniform, to simplify
the calculation.  Following the logic of Sec.~8.4.5 of \citet{BT}, we
find that the average collision time $t_c$ required for a planet to
pass within a distance $r_p$ from a star is given by
\begin{equation}\label{eq:tc}
t_c^{-1}=\frac{\nu\sqrt{\pi}}{2\sigma^3}\int_0^\infty
e^{-v_{\infty}^2/4\sigma^2} \left( v_{\infty}^3 r_p^2+2GM
  v_{\infty}r_p\right)\,dv_{\infty}. 
\end{equation}
Note that the latter term in parentheses 
above, representing gravitational focusing,
differs from that presented in Eq.~(8-121) of \citet{BT}, since they
are discussing the case of equal-mass stars, for which $M=2M_*$.
Here, for planet-star encounters, $M\simeq M_*$.
As we have seen in Fig.~\ref{fig:vcapt}, the condition for tidal
capture is more complicated than simply having the periastron fall
within a certain limit, since sufficient energy needs to be dissipated
to create a bound orbit.  Instead, for a given value of $v_\infty$,
the periastron separation must lie within a range of values
$r_1(v_\infty)\le r_p \le r_2(v_{\infty})$.  
We find that $r_1$ and $r_2$ can be determined implicitly, 
to more than sufficient accuracy, as roots of the relation
\begin{equation}\label{eq:vfit}
v_\infty=5\times 10^3 \frac{r-17.5}{(r/10.0)^{10}}~{\rm km/s},
\end{equation}
which is shown as a dot-dashed line on Fig.~\ref{fig:vcapt}.
We integrate Eq.~\ref{eq:tc} only up to the maximum possible capture
velocity, and only for ranges of periastron separations $r_1<r_p<r_2$,
for the three cases considered by \citet{Gaudi}: a globular cluster
($\sigma=10$~km/s, $\nu=10^4~{\rm pc}^{-3}$), a rich open cluster
($\sigma=1.5$~km/s, $\nu=10^3~{\rm pc}^{-3}$), and a loose stellar
association ($\sigma=0.6$~km/s, $\nu=10^2~{\rm pc}^{-3}$).  In all
cases, it is the gravitational focusing term which dominates, since
$GM_*/r_p\gg v_{capt}^2$ for each.
We find capture timescales, respectively, of $2.1\times 10^4$~Gyr (globular
cluster), $4.5\times 10^3$~Gyr (open cluster), and $1.3\times
10^4$~Gyr (loose association).  
The systematic error in estimating $v_{capt}$ for large values of
$r_p$, discussed above, plays essentially no role in our final results
for the globular cluster, and introduces an uncertainty of
approximately $2\%$ and $10\%$ in the rate for open clusters and
loose associations, respectively.  \citet{Gaudi} overestimates the
capture rate, due to a numerical error in the capture timescale
formula (Eq.~7; the denominator should have a 2, not an 8), and by
assuming capture for closer passages in which mass is lost but the
planet remains on a hyperbolic orbit.

Regardless of whether the planet's initial orbit is hyperbolic or
elliptical, we
expect that several distinct phenomena should occur for
passages with sufficiently small periastron separations.
For $r_p \lesssim 21$, we expect that
some amount of mass loss will occur, regardless of whether the system
ends up bound or unbound.  For all systems with $r_p \lesssim 17.5$, we
expect that no tidal capture can occur, since there is a net gain in
orbital energy.  In addition, we expect significant
mass loss ($\Delta M > 0.25 M$) for systems with $r_p \lesssim 15$.

The final orbital parameters for our runs are given in Table
\ref{table:1}. Note that mass
contained in the inner stream, $\Delta m_{in}$, is treated here as if it
will
eventually accrete onto the star when we determine the orbit.
While this may not hold in detail for all matter in the stream
(since some small fraction of the matter may be heated or shocked and
attain enough energy to be ejected form the system), its mass as a
fraction of the star's total mass is virtually negligible.  To compute
the orbital energy and angular momentum, we include both the
point-mass star and all particles bound to it when calculating its
center-of-mass, and only those particles bound to the star when
calculating its position.  These results are only intended to describe
the binary configuration present at the end of the dynamical encounter.
Thermal and radiative processes will certainly affect the structure of
the planet, and additional mass loss may occur, changing the orbital
parameters \citep{GS}.  Such phenomena cannot be properly described by an SPH
treatment, however.

\section{Conclusions}\label{sec:concl}

Based on recent observations of three extrasolar planets with orbital
periods of $< 2$ days but radii comparable to that of Jupiter, we
suggest that these planets may have either been scattered onto
very highly eccentric orbits through dynamical interactions in their
protoplanetary systems or tidally captured from free-floating
trajectories.  Through a series of tidal
encounters at periastron, energy can be dissipated until the orbit
eventually reaches the ideal circularization radius, 
at a distance almost exactly twice that of the
Roche limit.

Numerical calculations confirm that the classical 
Roche limit plays an important
role in these interactions, since it sets the boundary that
determines whether or not mass will be stripped from the planet during
periastron passage.  In turn, mass loss can play a significant role in
the resulting energy and angular momentum budgets, with regard to both
the orbital and spin evolution of the planet.

For orbits with periastron separations that lie outside of the Roche
limit, we find that a Press-Teukolsky expansion over 
linear, non-radial, adiabatic
oscillations provides a good estimate of the energy and angular
momentum transfer between the orbit and the planet.
For closer passages, however, the linear formalism breaks down, 
as mass is stripped off the planet through both the
inner (L1) and outer (L2) Lagrange points.  Still, we find that the
planet will never be fully disrupted on any orbit that does not
result in a collision with the star. 
Because the stripped mass acts as an energy sink, we find that
the maximum negative change in orbital energy for the planet occurs
for encounters just within the Roche limit.  For
smaller separations, $r_p \lesssim 19$, the energy loss from the
orbit decreases, since a greater fraction of the tidal energy is
used to unbind mass from the planet.  For periastron separations
$r_p\lesssim 17.5$, the net change in orbital energy turns
positive, and the planet, with smaller mass after the encounter, is
ejected from the system on a hyperbolic orbit.

Our results can be extended to predict the fate of earthlike planets
undergoing strong tidal encounters as well.  For the Earth-Sun system,
the Roche limit is reached at a periastron separation of $r_p=1.38
R_{\odot}$.  Strong tidal encounters, defined as those with
interaction strength $\eta =1$, only occur for passages with $r_p\le
0.64 R_{\odot}$, leading us to conclude that terrestrial planets would
have to pass well within the star in order to be tidally disrupted.  

The biggest problem for models which invoke long term tidal
circularization of orbits involves dissipating the energy injected
into the planet while the orbit 
circularizes.  Several planetary evolution calculations have
predicted that planets on elliptical orbits with periods 
$P\lesssim 3$ days should expand to the point where they fill their
Roche lobe, and begin to transfer mass to the parent star, eventually
leading to disruption of the planet \citep{GLB,GBL}.If this is true,
however, it is unclear how one can explain the past orbital evolution of
OGLE-TR-56b and the other systems whose orbital periods fall within
this limit, on what are now circular orbits. One of two possible explanations
seems to be required to explain the data we confront today.  First, if
planets are unable to survive the expansion resulting from tidal
heating during orbital circularization, we need to determine a
mechanism whereby planets migrate inward, perhaps during the evolution
of the protoplanetary disk, but stop at {\it twice the Roche limit}.
The answer seems not to lie in Roche lobe overflow for this case,
since recent observations \citep{Tor} and theoretical calculations
\citep{Burr} indicate that OGLE-TR-56b has a radius of $1.23\pm
0.16~R_\odot$, which is well within the Roche lobe.
Alternately, as presented here, these planets may undergo tidal
encounters with stars after being kicked into highly eccentric orbits
or captured from the field, eventually circularizing at a distance
equivalent to twice the Roche limit, as the current observed 
population suggests.  If this is true, then some new mechanism must be
constructed to explain how such planets radiate away the dissipated energy before
they are disrupted.  At the moment, there is no detailed theoretical
model which seems to provide the complete picture, but work is
underway to model the detailed evolution of rotating planets
undergoing a succession of tidal interactions during periastron
passages (see, e.g., \citealt{IP1}).

Since the change in orbital energy for the interactions we
investigated here is significantly larger than the total energy of the
systems, we conclude that our results should hold equally true for
nearly parabolic orbits with slightly positive total energy,
i.e. tidal captures of planets on weakly hyperbolic orbits with small
relative velocity $v_\infty$.  The parameter space for encounters that
lead to capture into a bound orbit is limited, however, primarily
because of the mass-stripping effect.  Essentially, if a planet passes
too far from the star, it will not be captured, but if it passes too
close, so much energy is used to unbind its outer layers that the
planet is boosted to an even higher energy hyperbolic orbit.  The
parameter space of of orbits leading to tidal capture is significantly greater for open
clusters, with a typical velocity dispersion of $\sim 2$ km/s, than
for globular clusters ($\sim 10$ km/s) but the dramatically
lower spatial density of planets and stars is likely to hinder the
capture process.  Still, based on approximate parameters for these
systems, although we expect that the capture timescale for planets will be
almost 5 times shorter in open clusters, the significantly greater
number of stars in globular clusters should lead to a higher overall
capture rate.

Our detailed calculations suggest that the results of \citet{Gaudi}
may need to be recalculated, since his assumptions about the fate of
planets after tidal interactions uses an overly simplistic dependence
on the periastron separation.  In that paper, he argues that according
to simple analytic approximations, a planet with mass $M_P=0.001 M_*$
must pass within two stellar radii ($r_p \lesssim 20 R_P$ for our
chosen parameters) in order to be captured.  We find that this upper limit
should be placed further out, but with a non-trivial dependence
between the initial relative velocity of the planet and star and the
critical periastron separation for capture.  We also find that a lower limit can
be placed on capturable orbits from the condition that the net change
in orbital energy must be negative, which only occurs for orbits with
$r_p\gtrsim 17.5$.  By adding these considerations to a proper
long-term treatment of the production and capture of free floating
planets in stellar clusters we may be able to pin down more accurately
the true capture rate and properties of the resultant systems.

\appendix

\section{The Roche Limit}\label{sec:roche}

Even though the duration of the close passage of a gas giant past
a star on an elliptical quasiparabolic orbit is extremely short in comparison to
the orbital period, it is relatively long compared to the internal hydrodynamical
timescale of the planet.  Thus, one might possibly guess that if the planet passes
within the classical Roche limit, it should lose mass, whereas those
that pass outside the Roche limit will not.  As we will demonstrate
below, this simple guess is manifestly correct, even though the
situation we consider is posed somewhat differently than the classic
Roche problem.

Our dynamical models violate three
assumptions that underlie the Roche lobe calculations.  First, planets
are extended objects, whereas the Roche lobe approximation is solved
for the gravitational potential field around a pair of point masses.
Second, the planets in our calculation are irrotational in the
inertial frame, rather than synchronized.  Third, the orbital velocity at
periastron passage is $\sqrt{2}$ times as large for a nearly parabolic
orbit as it is for a circular orbit, so the angular velocity of the
corotating frames are different.

\begin{figure}
\centering \leavevmode \epsfxsize=6in \epsfbox{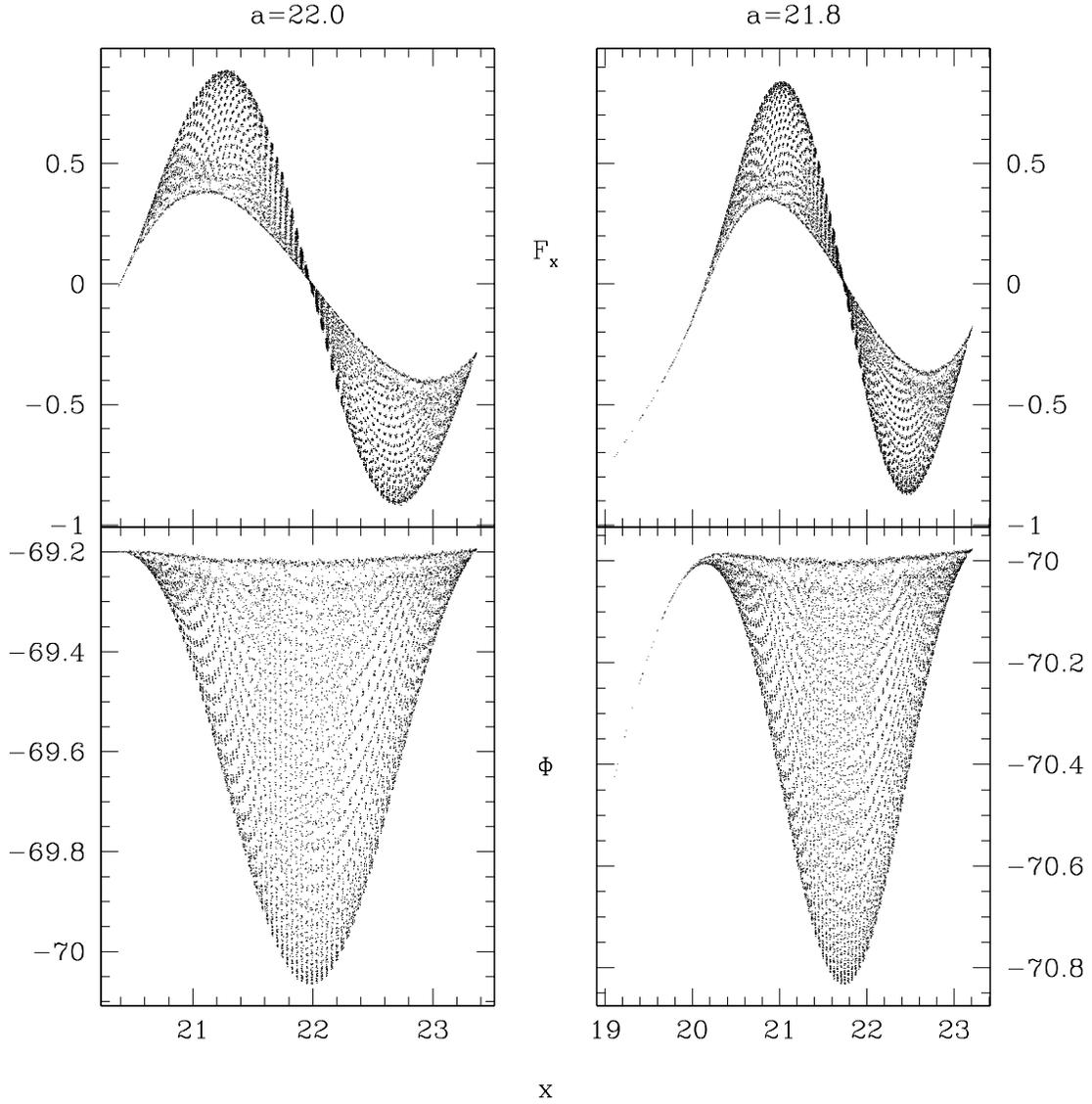}
\caption{Gravitational force $F_x$ along the orbital axis (top panels) and
  gravitational potential $\Phi$ (bottom panels) for synchronized quasi-equilibrium
configurations just outside the Roche limit at separation $a=22.0$
  (left panels) and just within the Roche limit at separation $a=21.8$
  (right panels).  For clarity, only particles near the equatorial
  plane, with $|z|<0.05$ are shown.  Our estimate of the Roche limit
  agrees extremely well with that from \protect\citet{Pac}.}
\label{fig:roche}
\end{figure}

It is relatively easy to test out the effects of the first of these
simplifications using SPH techniques.  Since the material in a synchronized
planet
on a circular orbit is stationary in the corotating frame, we can use
relaxation techniques to construct equilibrium configurations for a
given orbital separation.
First, to account for the rotating frame, we
add to the force equation a centrifugal acceleration term
$\vec{a}_{cent}=\Omega^2\vec{r}$, where the angular velocity $\Omega$
is calculated at every timestep so that the outwardly directed
centrifugal force exactly balances the inward gravitational force on
each member of the system.  To drive the system toward equilibrium, we
also add a velocity damping term,
$\vec{a}_{damp}=-\vec{v}/t_{relax}$, where we set $t_{relax}=1$.
The planet is initially laid down in a spherical configuration, with
an initial orbital separation $a_0=25$, slightly outside the Roche
limit, $a_R=21.64$, which we find for $q=0.001$ from
\citet{Pac}.
The planet is allowed to evolve for 25 dynamical times toward the proper
tidally
extended equilibrium state while the orbital separation is kept fixed.
Next, we slowly decrease the orbital separation, sufficiently slowly
that the planet can remain in quasi-equilibrium during the process.
Eventually, the planet fills its Roche lobe and matter crosses through
the inner Lagrange point toward the star.  As shown in
Fig.~\ref{fig:roche}, we find that this happens at a separation
$a\approx 22.0$, as depicted in the left hand panels.  In the top
plot, we show the gravitational force in the rotating frame in the
x-direction (the
direction of the orbital separation vector) as a function of position
for all particles near the orbital plane satisfying  $|z|<0.1$.
We see that particles on the innermost edge of the planet experience
almost no net gravitational force, indicating that they are extremely
close to the inner Lagrange point.  The bottom panel, showing the
gravitational potential for these same particles, with the centrifugal
barrier terms factored in, yields the same conclusion.  In the right
hand panels, we show the same quantities when the orbital separation
has been reduced to $a=21.8$, showing clear evidence that the
innermost particles making up the planet have crossed through the
inner Lagrange point and are now bound to the star instead.

Thus, we see that even though the exact conditions used to
determine the Roche limit do not apply in detail to extended
polytropic configurations, 
the simplest approximation formulae for the Roche lobe
radius as a function of the system mass ratio \citep{Pac,Egg} are
still accurate to within approximately $2\%$.
While the assumptions underlying the classical Roche limit, i.e., synchronized spins and the angular velocity
corresponding to a circular orbit, 
are violated by irrotational planets
on quasiparabolic orbits, we find that the critical point at which mass is
stripped from the planet does fall very near this line, and in a small
abuse of notation we will refer to the latter as the ``Roche limit''
as well.  It should be noted, though, that this latter quantity may
very well have weak but non-trivial dependence on the planet's spin
and the orbital parameters.

\section{Linear, non-radial, adiabatic oscillations}\label{sec:leeost}

So long as the periastron separation of the orbit is sufficiently
large, the perturbations to the planetary structure will remain
squarely in the linear regime.
Tidal capture through energy losses to adiabatic, non-radial
oscillations was studied in detail by \citet{PT},
who expanded greatly on the analytical treatment devised by
\citet{FPR}.  This work was extended by \citet{LO}, who determined
approximate power-law relations which govern the energy transfer
scaling at large separations, and corrected an error in the original
Press-Teukolsky paper.  Their work focused on $n=3/2$, $2$, and
$3$ polytropes, all of which had adiabatic indexes $\Gamma_1\equiv
d\ln P /d\ln \rho=5/3$,
appropriate for general polytropic stellar models.  
These results have been used to study a variety of phenomena related
to tidal capture, including the tidal damping of oscillations through
mode-mode coupling and luminosity variations for radio pulsars in
eccentric orbits (see, e.g., \citealt{KAQ,KG}).

Here, we extend
this method to cases appropriate for the planetary models we
calculate.  For gas giants, we assume an EOS with $n=1, \Gamma_1=2$,
and for an earthlike planet, we take $n=0.5, \Gamma_1=3.0$.  As
$n=1/(\Gamma_1-1)$ for these models, we know that we can ignore
g-mode oscillations, and are left only with the f-mode and p-mode
cases.

To determine the oscillation mode frequencies and overlap integrals,
corresponding to Tables 1A and 1B of \citet{LO}, we used a code which
solves four linked, linearized, first order equations of adiabatic motion
(see, e.g., \citealt{LW,Dzi}),
integrating inward from the surface and outward from the center, and
matching the solutions at the midpoint through relaxation techniques.
In table~\ref{table:qnl}, we list the squared oscillation
eigenfrequencies $\omega_n^2$ and overlap integrals $|Q_{nl}|$ for the
f-mode and p-modes up to $p_5$, for both $l=2$ and $l=3$ modes,
using the same conventions found in
\citet{LO}, which agree with those found here as well.

\begin{figure}
\centering \leavevmode \epsfxsize=6in \epsfbox{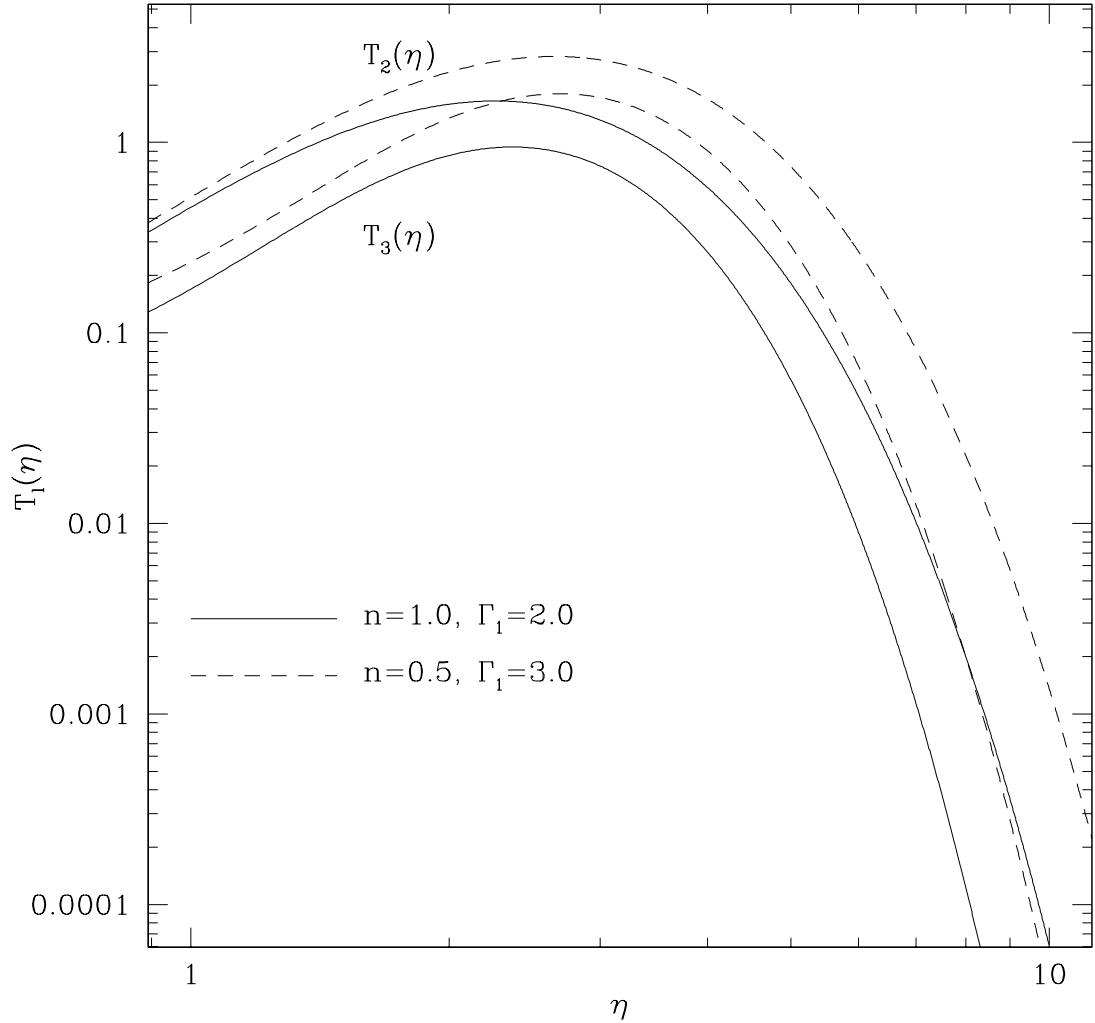}
\caption{Tidal energy transfer parameters $T_2(\eta)$
  and $T_3(\eta)$, as functions of the interaction strength $\eta$,
  defined by Eq.~\protect\ref{eq:eta}, for polytropic models with $n=1.0, \Gamma=2.0$
  (solid lines) and $n=0.5, \Gamma_1=3.0$ (dashed lines), computed
  from Eq.~(2.1) of \protect\citet{LO}. Note that in
  \protect\citet{LO}, the labels in Fig.~1a are reversed for the $l=2$
  and $l=3$ modes.}
\label{fig:tl}
\end{figure}

The tidal energy dissipation during an encounter can be expressed in
the parameterized form given by Eq.~(2.1) of \citet{LO}, noting that in
their first term, the superscript ``2'' was incorrectly transposed
outside the parentheses,
\begin{equation}
\Delta E_p = \left(GM_*^2/R_P\right) \sum_l
\left(\frac{R_P}{r_p}\right)^{2l+2}T_{l}(\eta),
\end{equation}
where the dimensionless separation parameter $\eta$ is defined by Eq.~\ref{eq:eta}.
Full details, including the derivation of the energy loss formula,
can be found in \citet{PT} and \citet{LO}.
In Fig.~\ref{fig:tl}, we show the dependence of $T_l(\eta)$ on $\eta$,
for $l=2$ and $l=3$ using both of the polytropic EOS discussed above.
We note in passing that these curves are qualitatively similar to the
$n=3/2$ case shown in Fig.~1a of \citet{LO}, who {\it reversed} the
labels for the $l=2$ and $l=3$ mode on their figure; in general, $l=2$
modes are almost always stronger for almost any physical system of
interest.  Since there are no low-frequency g-modes for these models,
we see the expected exponential drop-off at larger separations.  The
largest contribution to all the cases shown here comes from the
f-mode, which has the lowest frequency and is thus most coherently
driven by the relatively long period interaction.

\acknowledgements

This work was supported by NSF Grant AST-0206182.
FAR thanks the Kavli Institute for Theoretical Physics for hospitality and
support. 

\bibliography{planetbib}

\newpage
\begin{deluxetable}{c|ccccccccccc}
\tablewidth{6in}
\tabletypesize{\footnotesize}
\tablecaption{Run results \label{table:1}}
\tablehead{
\colhead{$r_p$} & \colhead{$\Delta m$} & \colhead{$\Delta m_{in}$} &
\colhead{$E_{orb}$} & \colhead{$J_{orb}$} &
\colhead{$r_{max}$} & \colhead{$E_{sp}$} & \colhead{$J_{sp}$} &
\colhead{$W$} & \colhead{$U$} & \colhead{$R_{95}$} & \colhead{$R_{100}$}}
\startdata
12 & 5.8E-1 & 3.2E-1 & 1.77E-1 & 62  & hyp & 4.68E-3 &
1.26E-1 & -0.052 & 0.014 & 8.77 & 57.2 \\
12.5 & 5.4E-1 & 3.9E-1 & 6.80E-2 & 70  & hyp & 6.66E-3 &
1.16E-1 & -0.077 & 0.021 & 7.68 & 62.8 \\
13 & 4.9E-1 & 2.7E-1 & 9.11E-2 & 81  & hyp & 9.94E-3 &
1.45E-1 & -0.103 & 0.027 & 6.75 & 58.2 \\
13.5 & 4.2E-1 & 2.4E-1 & 1.63E-1 & 93  & hyp & 1.30E-2 &
1.69E-1 & -0.140 & 0.038 & 5.87 & 57.1 \\
14 & 3.7E-1 & 2.1E-1 & 1.22E-1 & 105 & hyp & 1.53E-2 &
1.77E-1 & -0.177 & 0.049 & 5.35 & 56.3 \\
14.5 & 3.0E-1 & 1.8E-1 & 1.30E-2 & 117 & hyp & 1.72E-2 &
1.91E-1 & -0.224 & 0.063 & 5.06 & 61.1 \\
15 & 2.4E-1 & 1.5E-1 & 1.25E-1 & 131 & hyp & 2.27E-2 &
1.98E-1 & -0.279 & 0.077 & 3.58 & 38.8 \\
16 & 1.5E-1 & 9.3E-2 & 9.37E-3 & 153 & hyp & 2.36E-2 &
1.90E-1 & -0.385 & 0.111 & 2.75 & 36.1 \\
17 & 7.2E-2 & 5.0E-2 & -5.68E-2 & 171 & 1.63E3 & 2.30E-2 &
1.74E-1 & -0.491 & 0.147 & 1.82 & 37.4 \\
18 & 2.7E-2 & 2.1E-2 & -1.15E-1 & 184 & 8.42E3 & 1.97E-2 &
1.46E-1 & -0.577 & 0.177 & 1.30 & 37.6 \\
19 & 6.4E-3 & 5.7E-3 & -1.43E-1 & 193 & 6.95E3 & 1.37E-2 &
1.10E-1 & -0.637 & 0.202 & 1.11 & 38.4 \\
20 & 5.8E-4 & 5.6E-4 & -1.42E-1 & 200 & 7.04E3 & 7.60E-3 &
7.27E-2 & -0.676 & 0.218 & 1.02 & 38.2 \\
21 & 1.2E-6 & 3.5E-7 & -1.28E-1 & 205 & 7.82E3 & 3.45E-3 &
4.53E-2 & -0.700 & 0.229 & 0.97 & 33.9 \\
22 & 0.0 & 0.0 & -1.17E-1 & 209 & 8.49E3 & 1.35E-3 &
2.73E-2 & -0.715 & 0.235 & 0.94 & 1.11 \\
23 & 0.0 & 0.0 & -1.10E-1 & 214 & 9.06E3 & 5.19E-4 &
1.65E-2 & -0.726 & 0.240 & 0.93 & 1.06 \\
24 & 0.0 & 0.0 & -1.06E-1 & 219 & 9.40E3 & 1.92E-4 &
9.89E-3 & -0.733 & 0.242 & 0.92 & 1.03 \\
25 & 0.0 & 0.0 & -1.04E-1 & 223 & 9.60E3 & 7.71E-5 &
5.90E-3 & -0.740 & 0.246 & 0.91 & 1.03 \\
27 & 0.0 & 0.0 & -1.01E-1 & 232 & 9.89E3 & 1.48E-5 &
2.03E-3 & -0.745 & 0.248 & 0.90 & 1.00 \\
30 & 0.0 & 0.0 & -9.97E-2 & 244 & 1.00E4 & 1.85E-6 &
4.44E-4 & -0.747 & 0.248 & 0.89 & 0.98 \\
40 & 0.0 & 0.0 & -9.95E-2 & 282 & 1.00E4 & $<1$E-6 &
$<1$E-4 & -0.749 & 0.249 & 0.89 & 0.97 \\
50 & 0.0 & 0.0 & -9.95E-2 & 315 & 1.00E4 & $<1$E-6 &
$<1$E-4 & -0.749 & 0.249 & 0.89 & 0.97 \\
\enddata
\tablecomments{Results of our runs.  Here $r_p$ is the initial periastron
separation, $\Delta m$ the mass unbound from the planet, $\Delta m_{in}$
the amount of mass lost from the planet but bound to the star,
$E_{orb}$ and $J_{orb}$ the
final orbital energy and angular momentum, $r_{max}$ is
the new value of the apastron separation after the
encounter for systems which remain bound (``hyp'' indicates the planet
leaves on a hyperbolic orbit), $E_{sp}$ and $J_{sp}$ the
final spin energy and angular momentum of the planet, $W$ and $U$ the
planet's gravitational binding and internal heat energies, and
$R_{95}$ and $R_{100}$ are the radius of the final bound configuration
containing $95\%$ and all of the bound matter, respectively, with initial values of $0.89$ and $1.00$.  All energy
and angular momenta quantities are overall totals, not specific totals with the
mass dependence divided out.  Units are defined such that $G=M_P=R_P=1$.}
\end{deluxetable}

\begin{deluxetable}{c|cc|cc}
\tablewidth{4.0in}
\tablecaption{Squared oscillation eigenfrequencies and overlap integrals
\label{table:qnl}}
\tablehead{
\colhead{Mode} & \colhead{$(\omega^2)_{l=2}$} & \colhead{$(|Q_{nl}|)_{l=2}$}
& 
\colhead{$(\omega^2)_{l=3}$} & \colhead{$(|Q_{nl}|)_{l=3}$}}
\startdata
\tableline
& \multicolumn{4}{c}{$n=1.0, \Gamma_1=2.0$} \\
\tableline
$f$ & 1.505 & 5.558E-1 & 2.884 & 5.845E-1 \\
$p_1$ & 11.98 & 2.689E-2 & 15.79 & 4.053E-2 \\
$p_2$ & 29.32 & 2.610E-3 & 35.61 & 4.294E-3 \\
$p_3$ & 52.68 & 3.128E-4 & 61.56 & 5.352E-4 \\
$p_4$ & 81.85 & 4.007E-5 & 93.38 & 7.113E-5 \\
$p_5$ & 116.7 & 4.699E-6 & 130.9 & 9.319E-6 \\
\tableline
& \multicolumn{4}{c}{$n=0.5, \Gamma_1=3.0$}\\
\tableline
$f$ & 1.097 & 6.236E-1 & 2.230 & 7.093E-1 \\
$p_1$ & 18.50 & 6.613E-3 & 24.14 & 1.178E-2 \\
$p_2$ & 47.36 & 5.895E-4 & 57.64 & 5.594E-4 \\
$p_3$ & 86.33 & 3.780E-4 & 101.3 & 3.138E-4 \\
$p_4$ & 135.2 & 3.072E-4 & 155.0 & 2.669E-4 \\
$p_5$ & 193.9 & 2.569E-4 & 218.5 & 2.254E-4 \\

\enddata
\tablecomments{Dimensionless 
squared oscillation eigenfrequencies $\omega^2$ and
  overlap integrals $|Q_{nl}|$ for the $l=2$ and $l=3$ modes of
    polytropic models, defined in Eqs.~2.3--2.7 of \protect\citet{LO}.  Units are such that $G=M_P=R_P=1$, as used throughout.}
\end{deluxetable}

\end{document}